\documentclass[twocolumn,unsortedaddress,aps,prl]{revtex4}
\usepackage{graphicx}
\usepackage{amsmath}

\begin{document}

\title{Spin chirality on a two-dimensional frustrated lattice}

\author{Daniel Grohol,$^{1}$ Kittiwit Matan,$^{2}$ Jin-Hyung Cho,$^{2}$
Seung-Hun Lee,$^3$ Jeffrey W. Lynn,$^3$ Daniel G. Nocera,$^{1}$ and
Young S. Lee,$^{2*}$}

\affiliation{
$^1$Department of Chemistry, Massachusetts Institute
of Technology, Cambridge, Massachusetts 02139, USA\\
$^2$Department of Physics and Center for Materials Science and
Engineering, Massachusetts Institute of Technology, Cambridge,
Massachusetts 02139, USA\\
$^3$NIST Center for Neutron Research, Gaithersburg, Maryland
20899, USA \\
$^*$e-mail:younglee@mit.edu
}


\begin{abstract}

The collective behavior of interacting magnetic moments can be
strongly influenced by the topology of the underlying lattice.  In
geometrically frustrated spin systems, interesting chiral
correlations may develop that are related to the spin arrangement
on triangular plaquettes.  We report a study of the spin chirality
on a two-dimensional geometrically frustrated lattice.  Our new
chemical synthesis methods allow us to produce large single
crystal samples of KFe$_3$(OH)$_6$(SO$_4$)$_2$, an ideal
Kagom\'{e} lattice antiferromagnet.  Combined thermodynamic and
neutron scattering measurements reveal that the phase transition
to the ordered ground-state is unusual. At low temperatures,
application of a magnetic field induces a transition between
states with different non-trivial spin-textures.
\end{abstract}

\maketitle

Geometrically frustrated magnets are unusual in that they may have
disordered ground-states in which an enormous number of spin
configurations share the same energy.\cite{Ramirez,Gingras1} The
Kagom\'{e} lattice antiferromagnet, formed of corner sharing
triangles, is one of the most highly frustrated two-dimensional (2D)
lattices.  Though long regarded as a prime model for studying spin
frustration,~\cite{Ramirez,Harris,Sachdev,Chalker1,Ritchey} the
Kagom\'{e} lattice has escaped precise magnetic characterization
because compounds comprising this lattice are difficult to make pure
and in large single crystal form.  We present a study of an ideal
Kagom\'{e} lattice compound, the iron jarosite ${\rm
KFe_3(OH)_6(SO_4)_2}$.  This material possesses robust chiral
correlations related to the arrangement of spins around triangular
plaquettes.~\cite{Inami,Wills2}  The presence of spin chirality in
condensed matter systems may play a role in important phenomena
ranging from high-temperature (high-$T_C$)
superconductivity~\cite{Wen} to the anomalous Hall
effect~\cite{Nagaosa1,Nagaosa2}.  Currently, there are relatively
few experimental studies of spin chirality in frustrated magnets.
Here, we study the non-trivial spin-textures related to chirality in
a jarosite material in both the spin-ordered and spin-disordered
states. Our magnetic susceptibility, specific heat, and magnetic
neutron scattering measurements on pure single crystals show that
the transition to the low temperature ordered phase is unusual.  In
addition, we find that the spin-texture of the ground state can be
controlled by applying a magnetic field.

\begin{figure}
\includegraphics[width=3.0in]{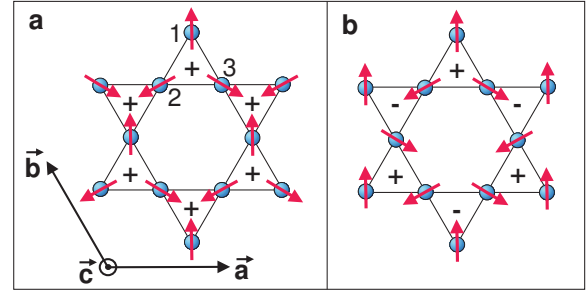}
\vspace{0mm} \caption{The Kagom\'{e} lattice with spins arranged in
two different configurations.  \textbf{a}, The ``$q$=0'' structure,
which is the ground state configuration for iron jarosite. The spin
arrangement has uniform, positive vector chirality, indicated by the
+ within each triangular plaquette.  \textbf{b}, An alternate spin
arrangement with staggered vector chirality, known as the
``$\sqrt{3} \times \sqrt{3}$'' structure.}
\end{figure}

The magnetic Fe$^{3+}$ ions of jarosite have spin-5/2 and lie at the
corners of the triangles of the Kagom\'{e} lattice, shown in Figure
1. To first approximation, the spin Hamiltonian is given by
\begin{equation}
H=\sum_{\langle i,j \rangle} J \vec{S}_i \cdot \vec{S}_j\nonumber
\end{equation}
where $J$ is the nearest neighbor magnetic exchange interaction
($J>0$). Because the interaction is antiferromagnetic, the spin
system is frustrated, and the corner-sharing arrangement leads to a
higher degree of degeneracy for the ground state than the triangular
lattice. The iron jarosite is a particularly ideal Kagom\'{e}
lattice compound for the following reasons. First, it consists of
single layers of undistorted Kagom\'{e} planes, and these planes
remain undistorted down to the base temperatures of our measurements
($T\sim 5$~K). Second, this jarosite can be synthesized with
compositions that are stoichiometrically pure, as we describe below.
This ensures that we are primarily studying the effects of geometric
frustration rather than the effects of disorder. Third, large single
crystals can be made, which allow investigation of the spin
correlations of this Kagom\'{e} compound that would not be possible
with powder samples alone.

Until recently, jarosites have been prepared typically by
precipitation under hydrothermal conditions:
\begin{align}
3{\rm \;Fe^{3+}~+~} & {\rm 2\;K_2SO_4~+~6\;H_2O} \rightarrow \nonumber\\
 & {\rm KFe_3(OH)_6(SO_4)_2~+~3\;K^+~+~6\;H^+}. &
\end{align}
Under these conditions, the monovalent K$^+$ cation is susceptible
to replacement by hydronium ions and the coverage of the Fe$^{3+}$
lattice sites is incomplete.  Also, only microcrystalline materials
are obtained owing to the heightened acidity of the solution as well
as the speed and intractability of the precipitation
reaction.\cite{Wills1,Inami} The challenges confronting the
synthesis of pure jarosites have been overcome with the development
of redox-based hydrothermal methods.~\cite{Nocera}  Control over the
precipitation of the jarosite is achieved by inserting two
oxidation-reduction steps before reaction (1)
\begin{align}
{\rm Fe + 2\;H^+} & \rightarrow {\rm Fe^{2+} + H_2} \\
{\rm 2\;Fe^{2+} + \frac{1}{2}\;O_2  + 2\;H^+} & \rightarrow {\rm
2\;Fe^{3+} + H_2O}.
\end{align}
In this manner, the Fe$^{3+}$ is slowly generated throughout the
course of the hydrothermal process and the pH is moderated because
three overall equivalents of protons are consumed in the production
of an equivalent of jarosite.  We have refined the synthesis process
to optimize the size of the single crystals.  Crystals as large as
10 mm in length and 48 mg in mass have been grown, making possible
the inelastic neutron scattering measurements that we discuss below.
Details of the single crystal synthesis may be found in the
Supplementary Information.

Even though the Kagom\'{e} lattice antiferromagnet should not order
at any non-zero temperature, powder neutron diffraction measurements
on ${\rm KFe_3(OH)_6(SO_4)_2}$ indicate that the spins order in a
coplanar `$q$=0' arrangement below $T_N \sim
65$~K.\cite{Inami,Wills2} In the $q$=0 structure, the spins on each
triangle are oriented at $120^\circ$ to each other, and the 2D
magnetic unit cell is identical to the 2D structure unit cell, as
shown in Fig.~1a. The ordered spins can be decomposed into three
sublattices, with the spins on each triangle labeled as $\vec{S_1}$,
$\vec{S_2}$, and $\vec{S_3}$.  The {\em vector chirality} for each
triangle may be defined as:
\begin{equation}
\vec{K}_V = \frac{2}{3\sqrt{3}}
(\hat{S_1}\times\hat{S_2}~+~\hat{S_2}\times\hat{S_3}~+~\hat{S_3}\times\hat{S_1}).
\end{equation}
For the coplanar arrangement, this vector is parallel to the
$c$-axis with amplitude +1 or -1.~\cite{Inami}  The neutron powder
results\cite{Inami,Wills2} indicate that each triangle has
positive chirality (+1) in the ordered state, such that the spins
point directly toward or away from the center of each triangle.

\begin{figure}
\includegraphics[width=3.0in]{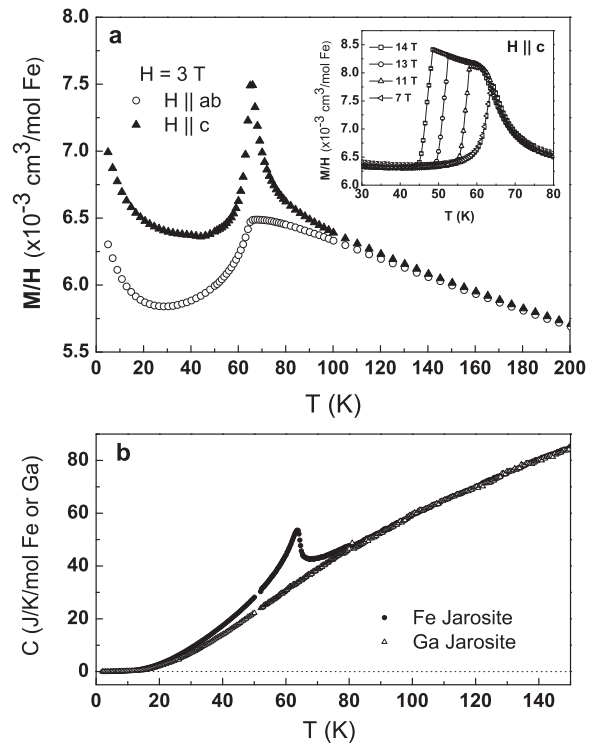}
\vspace{0mm} \caption{Magnetization and specific heat measurements
of ${\rm KFe_3(OH)_6(SO_4)_2}$.  \textbf{a}, $M/H$ versus
temperature with the applied field in two orientations, $H
\parallel c$ and $H \parallel ab$ measured on a single crystal sample
using a SQUID magnetometer. The inset shows $M/H$ for $H
\parallel c$ for fields up to 14 Tesla measured using an AC coil
set. The upturn in the magnetization at low temperatures (below 20
K) can be described by Curie behavior in which the density of free
spins corresponds to $\sim0.3\%$ of the total number of spins.
\textbf{b}, Specific heat of powder samples of ${\rm
KFe_3(OH)_6(SO_4)_2}$ (closed symbols) and the non-magnetic
isostructural compound ${\rm KGa_3(OH)_6(SO_4)_2}$ (open symbols).
The ${\rm KGa_3(OH)_6(SO_4)_2}$ data are scaled to match the
Fe-jarosite data at high temperatures and are used to estimate the
phonon contribution.}
\end{figure}

For this compound, like as for real magnetic materials, the
interaction Hamiltonian of the spins contains terms beyond isotropic
Heisenberg exchange.  These additional terms cause the system to
order at a non-zero $T_N$ and determine the ground-state spin
arrangement. Currently, it is not clear whether the transition to
long-range order is driven by weak interplanar coupling,
spin-anisotropy, anisotropic exchange or some combination of these.
An ordered phase on a Kagom\'{e} lattice is characterized by two
order parameters (the sublattice magnetization and the vector
chirality) which have different symmetries.  An intriguing
possibility is that these symmetries are broken at different
temperatures. This has been proposed as a result of numerical work
on $XY$ or planar triangular lattice
systems.~\cite{Kawamura,Calabrese} However, this has not been
conclusively observed by experiment.~\cite{Mason,Plakhty} For
Kagom\'{e} lattice systems, relatively little is known about the
nature of the phase transitions that occur.

\

\noindent {\bf Magnetization and field-induced transition}

\

To address the above issues, we have first measured the
magnetization of a single crystal sample (mass of 13.5 mg) of ${\rm
KFe_3(OH)_6(SO_4)_2}$ using a SQUID magnetometer. Measurements were
taken with the applied field oriented along the $c$-axis ($H
\parallel c$) and within the $ab$ plane ($H
\parallel ab$) as shown in Fig.~2a. At low fields ($H<5$~T) with
the field along $c$, a sharp peak appears near 65 K, indicative of
the transition to the 3D magnetically ordered state.  This is
consistent with previous measurements on powder samples prepared
under similar synthesis conditions.~\cite{Grohol} When the field is
aligned along the $ab$-direction, the sharp peak is absent and is
replaced by a broad cusp.

In Figure~2b, we plot the specific heat $C$ of powder samples of
${\rm KFe_3(OH)_6(SO_4)_2}$.  A peak in the specific heat is found
at the magnetic transition temperature of 65~K. The entropy
associated with the 3D magnetic transition (integrating $C/T$ over
the temperature range from 2~K to 100~K) represents $\sim50$\% of
the $Rln6$ (where $R$ is the molar gas constant) total entropy
expected for the spin 5/2 system. This suggests that short-range
correlations have already formed at much higher temperatures.

At high temperatures ($T>150$~K), the susceptibility is isotropic
and follows a Curie-Weiss law $\chi = C/(T-\theta_{CW})$, consistent
with previous results on powder samples.~\cite{Grohol} Fits to this
law between 150~K and 550~K yield the values $\theta_{CW} =
-800(30)$~K, and $C = 5.6(2)$~cm$^3$ K/mol Fe. Because the data are
taken for $T < |\theta_{CW}|$, we extract the effective moment
$\mu_{\rm eff}$ and the nearest neighbor exchange coupling $J$ using
the high-temperature series analysis of Harris {\em et.al.}~for the
Kagom\'{e} lattice.~\cite{Harris}  Our results indicate that $J =
45(2)~{\rm K} = 3.9(2)~{\rm meV}$ and $\mu_{\rm eff} = 6.3(2)~\mu_B$
(close to the spin-only value of $5.92~\mu_B$ for Fe$^{3+}$).  We
note that a small next-nearest neighbor interaction ($J_{2} > 0$)
would serve to reduce the calculated value of $J$ by an amount of
order $J_{2}$.

The peak in $M/H$ for $H \parallel c$ at $T=65$~K indicates the
presence of weak ferromagnetism along the $c$-direction. Recent
theoretical work shows that antisymmetric exchange, via the
Dzyaloshinsky-Moriya (DM) interaction, may induce such a moment by
canting the spins slightly out of the plane.\cite{Elhajal}  This
interaction is present if there is no inversion center between
magnetic ions and adds the term
\begin{equation}
\sum_{\langle i,j \rangle} \vec{D}\cdot\left(\vec{S}_i \times
\vec{S}_j\right)\nonumber
\end{equation}
to the spin Hamiltonian, where $\vec{D}$ is the DM
vector.~\cite{Moriya}  This interaction causes the spins on each
triangle in the jarosite to form an `umbrella' structure and gives
each Kagom\'{e} plane a net ferromagnetic (FM) moment. However, the
interlayer coupling in iron jarosite causes the FM moments to couple
antiferromagnetically between layers in the absence of an applied
field, as shown in the inset of Fig.~3b. Our measurements on single
crystals allow us to explore this model of canted moments. The inset
of Fig.~2a shows $M/H$ as a function of temperature measured in high
fields ($H \geq 7$~T). At these high fields, the peak broadens and
the downturn in the magnetization shifts to lower temperatures. Such
behavior has been observed in the square-lattice antiferromagnet
La$_2$CuO$_{4}$~\cite{Wells} for which the DM interaction does
indeed result in weak ferromagnetism.\cite{Thio}

To investigate this spin canting further, we made magnetization
measurements as a function of magnetic field along $c$ direction as
shown in Fig.~3a. The results show an abrupt change in the
magnetization at a critical field, $H_C$, which we define as the
field at which $dM/dH$ is a maximum.  We interpret this abrupt
increase as a change from canted moments being oppositely directed
between planes to canted moments aligned in the same direction. This
is most likely caused by a $180^\circ$ rotation of all spins on the
layers that were previously oppositely canted, as shown in the inset
of Fig.~3b. The critical field as a function of temperature is also
shown in Fig.~3b. For comparison, we plot the integrated intensity
of the (1,1,$\frac{3}{2}$) magnetic Bragg peak measured with neutron
diffraction in Fig.~3c.  We find that $H_C$ scales quite closely
with the staggered moment $M^\dagger$ (which is proportional to the
square-root of the Bragg intensity). The mean-field result for
La$_2$CuO$_4$ gives $H_C \propto M^\dagger / \chi^\dagger$, where
$\chi^\dagger$ is the 2D staggered susceptibility.\cite{Thio} In our
case, the staggered susceptibility for the Kagom\'{e} lattice is
expected a have weaker temperature-dependence than the square
lattice due to the geometrical frustration.~\cite{Reimers}

\begin{figure}
\includegraphics[width=3.0in]{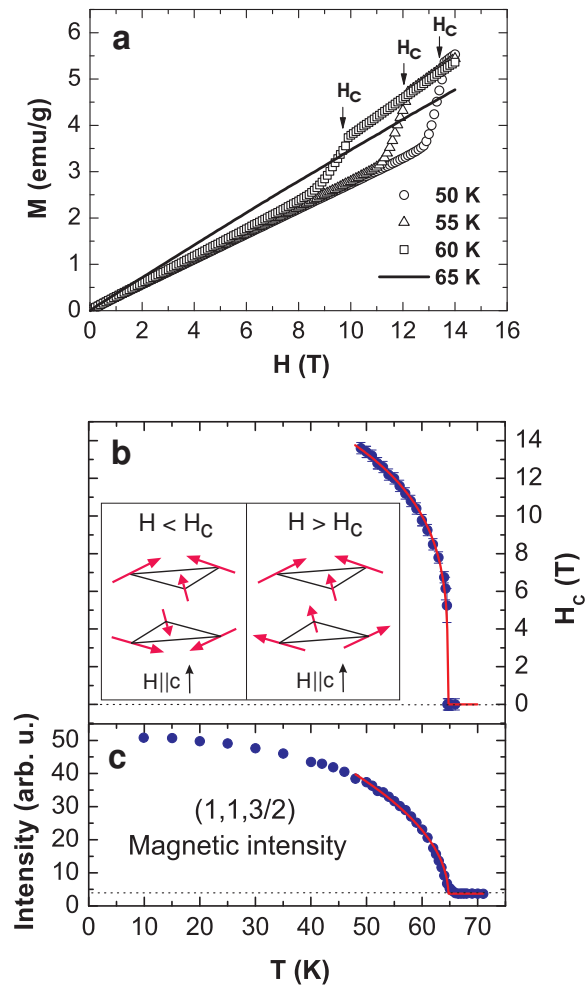}
\vspace{0mm} \caption{Measurements of the field-induced transition
to a state with non-zero scalar chirality . \textbf{a},
Magnetization versus applied field for several temperatures with $H
\parallel c$.  The $H_C$ values indicate the fields where $dM/dH$ is
maximum. \textbf{b}, $H_C$ versus temperature. The inset depicts the
change in the spin configuration below and above $H_C$. \textbf{c},
Integrated intensity of the (1,1,1.5) magnetic Bragg peak measured
with neutron diffraction on a single crystal in zero field.  This
quantity is proportional to $(M^\dagger)^2$, the square of the
staggered moment. The lines in panels B and C correspond to
power-law forms for $H_C$ and $M^\dagger$ with the exponent
$\beta=0.25$ and $T_N=64.7$.}
\end{figure}

Below $T=49$~K, the critical field becomes larger than 14 T (the
maximum field of our magnetometer).  Hence, it is difficult to
extract the zero-temperature values for $H_C$ and the canted
moment. However, an estimate for the canted moment can be made by
considering the jump in the magnetization in the vicinity of
$H_C$. For the data taken at $T=50$~K, we obtain a canting angle
for the ordered moment of $0.65(6)^\circ$ with respect to the
Kagom\'{e} plane for $H \simeq H_C$ (this angle will increase upon
cooling, as the order parameter has not yet reached its low
temperature value). The field-induced transition results from a
competition between the interlayer coupling $J_{\perp}$ and the
Zeeman energy; the magnitude of $J_{\perp}$ may be estimated from
the following relation: $H_C(0) M_F(0) = 2 S^2 |J_{\perp}|$, where
$H_C(0)$ and $M_F(0)$ are the critical field and FM moment per Fe
atom, respectively, at $T=0$. Our results indicate $|J_{\perp}| =
0.007(6)$~meV where the large error bar comes from the uncertainty
in extrapolating $H_C$ and $M_F$ to $T=0$. We find that the
magnitude of $J_{\perp}$ is several hundred times smaller than the
nearest-neighbor $J$, attesting to the two-dimensionality of the
system.

The field-induced spin-canting transition corresponds to a
non-trivial change in the spin-texture of the jarosite sample. In
particular, the transition yields a net, non-zero value for the
{\em scalar chirality}, defined on each triangular plaquette as
\begin{equation}
K_S = \vec{S_1}\cdot (\vec{S_2} \times \vec{S_3}).
\end{equation}
The presence of this type of chirality (in static or fluctuating
forms) can have important consequences in strongly correlated
electron systems, such as yielding an anomalous Hall effect in
metallic materials.~\cite{Nagaosa1,Nagaosa2} For $H<H_C$ the net
scalar chirality for our jarosite sample is zero because the
contributions from neighboring planes are equal and opposite.
However, for $H>H_C$ the spins flip (as depicted in Fig.~3b), and
the net scalar chirality becomes non-zero. There are few materials
with non-zero scalar chirality in the ordered state, especially on a
two-dimensional lattice. In iron jarosite, we have discovered a
phase transition in which a net scalar chirality can be `switched
on' by a magnetic field.

\

\noindent {\bf Neutron scattering and spin fluctuations}

\

To probe the microscopic behavior of the magnetism in the
temperature regime above $T_N$, we made inelastic neutron scattering
measurements of the spin fluctuations at $T=70$~K within the $L=0$
plane of reciprocal space.  For the temperature range $T_N < T <
120$~K, the susceptibilities for $H
\parallel c$ and $H \parallel ab$ deviate from each other, indicative
of the growing influence of spin-anisotropy.  For the 3D ordered
spin structure below $T_N$, the stacking arrangement of the planes
doubles the magnetic unit cell with respect to the structural unit
cell such that the magnetic Bragg peaks occur at half-odd-integer
values of $L$ ($L \neq 0$). However, for temperatures above $T_N$,
the correlations between layers are destroyed, and the 2D spin
fluctuations yield ``rods'' of scattering in reciprocal space along
the $L$ direction.  We have verified that all of the signal in
Fig.~4a disappears upon cooling below $T_N$, as expected for the
transfer of the intensity of the 2D critical scattering into 3D
Bragg points lying out of the scattering plane. Therefore, the scans
in Fig.~4a pass through these 2D rods of scattering and directly
measure the dynamic structure factor of the spin correlations of the
single Kagom\'{e} planes in iron jarosite. The instantaneous spin
correlation length, measured separately in an energy-integrating
configuration, is $\xi = 20(2)$~\AA~at this temperature.
\begin{figure*}
\includegraphics[width=6.5in]{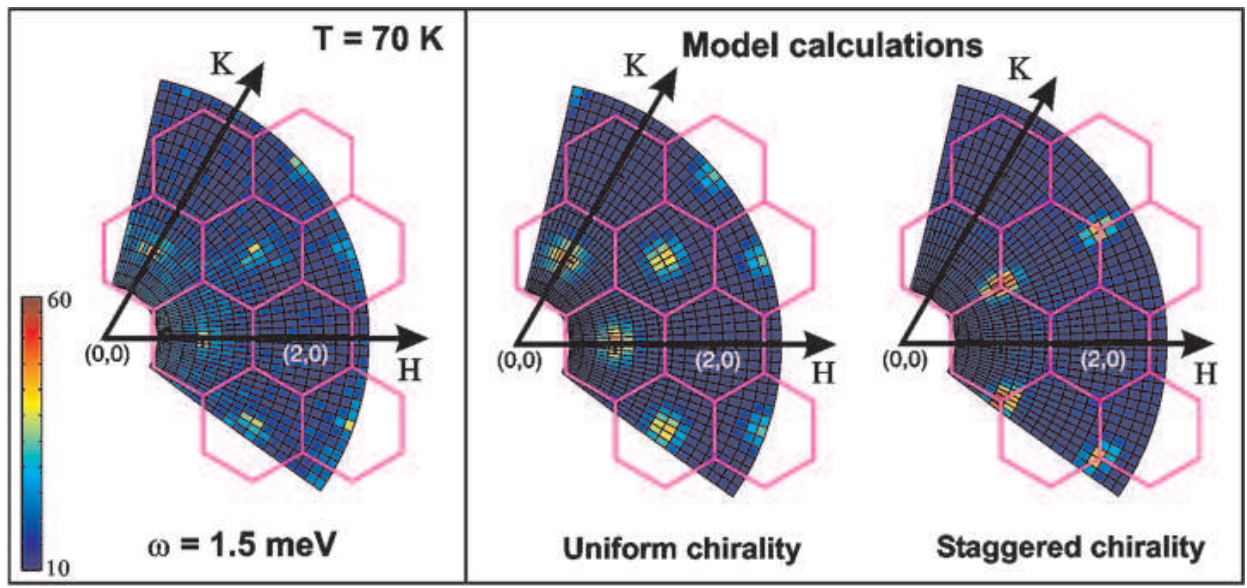}
\vspace{5mm} \caption{Inelastic neutron scattering data for ${\rm
KFe_3(OH)_6(SO_4)_2}$ measured above $T_N$, along with structure
factor calculations.  \textbf{a}, Intensity contour plot of data
from inelastic neutron scattering measurements of a single crystal
sample (mass=48 mg). The bright regions are the loci of scattering
intensity for the low energy ($\omega=1.5$~meV) spin fluctuations at
$T=70$~K above $T_N$.  \textbf{b}, Model calculations of the
intensity as described in the text.  The left plot corresponds to
short-ranged $q=0$ correlations with uniform vector chirality,
whereas the plot on the right depicts $\sqrt{3} \times \sqrt{3}$
correlations. Both calculations are for the case in which there is
no preferred spin direction within the Kagom\'{e} plane.}
\end{figure*}
A reciprocal space map of the intensity of the spin fluctuations is
shown in Figure~4a.  Outlined in red are the boundaries of the
structural Brillioun zones (BZ).  The strongest scattering occurs at
the centers of certain BZ.  The neutron scattering intensity can be
calculated by
\begin{equation}
I \; \propto \; f(\vec{Q})^2 \: \sum_{\alpha,\beta} \;
(\delta_{\alpha \beta} - \hat{Q}_\alpha \hat{Q}_\beta) \;
S^{\alpha \beta}(\vec{Q}, \omega)
\end{equation}
where $\alpha,\beta$ refer to $x,y,z$ vector components,
$f(\vec{Q})$ is the magnetic form factor, and $S(\vec{Q}, \omega)$
is the space- and time- Fourier transform of the spin-spin
correlation function.  The intensity variation reveals a wealth of
information about the short-range ordered state in the critical
regime just above $T_N$.

First, the fact that the intensities are centered within the
structural Brillioun zones indicates that the fluctuations have
the $q$=0 arrangement.  The absence of scattering at the (2,0,0)
position is consistent with this spin arrangement. Second, the
nearly equivalent intensities at the (1,0,0) and (1,1,0) positions
reveal that the spin direction on each sublattice is not fixed.
Rather, the two orthogonal spin directions within the Kagom\'{e}
plane are equally likely.  Hence the spins fluctuations have $XY$
symmetry at this temperature (the spin-rotational symmetry within
the Kagom\'{e} plane is not broken). In contrast, at low
temperatures in the 3D ordered state, a preferred spin direction
is chosen, and the intensities at (1,0,$L$) and (1,1,$L$) differ
considerably.~\cite{Inami,Nishiyama}

The observation of short-range $q$=0 correlations in the 2D
fluctuations implies a particular arrangement of the {\em vector}
chirality. In each region of correlated spins, the vector chirality
must be uniform (all positive or all negative for each plaquette).
Using equation (6), we calculated the neutron scattering intensity
arising from $q$=0 correlations between coplanar spins assuming only
positive chirality. Our model calculation are based on 7-unit cell
clusters of spins, and we have averaged over all spin directions
within the Kagom\'{e} plane. The results are shown in Figure~4b, and
the agreement with the data is excellent. Also shown in Fig.~4b is a
calculation for the intensity if the chirality were staggered as in
the $\sqrt{3} \times \sqrt{3}$ arrangement. If such correlated
regions exist, the error bars on our data indicate that the fraction
must be less than 5\%.

Our results shed light on several basic questions regarding the
magnetic phase transition in iron jarosite. First, we find that the
instantaneous spin correlations above $T_N$ are two-dimensional in
nature, and the $q$=0 arrangement is preferred. Therefore, the
selection of $q$=0 order (as opposed to $\sqrt{3} \times \sqrt{3}$
order which is predicted to be the preferred ground state for the
pure Heisenberg model ~\cite{Sachdev,Chubukov}) is caused by
interactions within a single Kagom\'{e} layer and is not controlled
by the interplanar interaction~\cite{Lee_Cr}. Second, our neutron
measurements reveal critical spin fluctuations above $T_N$ that have
$XY$ symmetry; hence, the magnetic ordering is not driven by 2D
Ising physics~\cite{Inami}. Most interestingly, we find that the
spin-rotational symmetry and the vector chiral symmetry are not
broken simultaneously at $T_N$.

The presence of vector chiral order above $T_N$ may be naturally
explained in light of the DM interaction.  One possibility is that
the DM interaction is the dominant source of spin anisotropy in iron
jarosite.  In this case, the vector chiral order appears
concomitantly with the growing spin correlations and does not
represent a spontaneously broken symmetry.  Another possibility is
that the $XY$ anisotropy has an origin (such as symmetric exchange
anisotropy~\cite{Yildirim}) distinct from the DM interaction. Once
the spin correlations become coplanar within the Kagom\'{e} plane,
the vector chirality of a triangular plaquette becomes a discrete
symmetry (the chirality vector is either up or down with respect to
the $c$-axis).  Hence, long-range chiral order is not precluded by
the Mermin-Wagner theorem.~\cite{Mermin} However, chiral order is
easily disrupted on a Kagom\'{e} lattice by the proliferation of
domain walls (thermally induced defects) which can form with little
cost in energy.\cite{Reimers,Henley,Chalker1} Then the presence of a
small DM interaction (in particular, the non-zero out-of-plane
component $D_z$) selects a particular chirality for each triangle,
and thereby inhibits domain wall formation. It remains possible that
the vector chirality represents a spontaneously broken symmetry;
however, clarification of this point requires further neutron
scattering measurements at higher temperatures.

Our measurements of this ideal Kagom\'{e} material reveal new
magnetic behavior related to two types of spin chirality: vector and
scalar. The DM interaction is a significant perturbation to the
Heisenberg Hamiltonian and strongly influences the low temperature
physics. For $T>T_N$, the vector spin chirality is ordered even in
the absence of broken spin-rotational symmetry. In the ordered state
below $T_N$, we have discovered a field-induced transition to a
state with non-zero scalar chirality.  Thus, materials based on
jarosites may be promising candidates for studies of the coupling
between non-trivial spin textures and the transport of electrons in
frustrated systems.  For example, carrier-doped compounds would
probably show an anomalous Hall effect of topological
origin,~\cite{Nagaosa1} and this might have useful applications in
spin-based electronics.

\

\noindent {\bf Methods}

The neutron scattering data were taken at the NIST Center for
Neutron Research. The integrated intensity of the (1, 1, 3/2)
magnetic Bragg peak (Figure 3c) was measured using the BT7
spectrometer. The incident neutron energy was fixed at 13.46 meV and
the horizontal collimation sequence was
open-open-sample-40$^\prime$-open.  The (002) reflection of a
pyrolytic graphite (PG) crystal was used as an analyzer.  For the
two-axis (energy-integrating) measurement of the instantaneous spin
correlation length at T=70 K, we used the BT9 spectrometer with the
sample aligned in the (HK0) scattering zone. The incident neutron
energy was 35 meV, with a collimation sequence of
40$^\prime$-24$^\prime$-sample-20$^\prime$. In both measurements, a
PG filter was placed in the incident beam to remove higher order
neutrons.  The inelastic neutron scattering data shown in Figure 4
were taken using the SPINS NG5 triple-axis spectrometer. The final
neutron energy was fixed at 5~meV and the horizontal collimation
sequence was ${ \; guide-80^\prime-S-40^\prime (radial)-open}$.  A
liquid-nitrogen-cooled beryllium filter was placed in the scattered
beam to remove higher order neutrons.

\

\noindent {\bf Acknowledgements}

We thank A. Clearfield for providing the large hydrothermal vessels
used in the crystal growth.  We thank Y.-B. Kim, T. Senthil, and T.
Yildirim for useful discussions.  This work was supported in part by
the MRSEC Program of the National Science Foundation under award
number DMR 02-13282 and also by NSF award DMR-0239377.

Correspondence should be addressed to Y.S.L.

Supplementary Information accompanies the paper on
www.nature.com/naturematerials.

\bibliography{kagome}
\bibliographystyle{Nat_mat}

\end{document}